\documentclass[fleqn,usenatbib,useAMS]{mnras}

\usepackage[T1]{fontenc}
\usepackage{ae,aecompl}

\usepackage{graphicx}	
\usepackage{amsmath}	
\usepackage{amssymb}	

\usepackage{multicol}       
\usepackage{bm}              
\usepackage{pdflscape}   

\usepackage{verbatim}
\usepackage{color}
\usepackage[usenames,dvipsnames]{xcolor}
\definecolor{green}{rgb}{0.0, 0.4, 0.0}
\definecolor{forestgreen(web)}{rgb}{0.13, 0.55, 0.13}
\definecolor{green(web)}{rgb}{0.13, 0.55, 0.13}
\definecolor{green}{rgb}{0.0, 0.4, 0.0}

\title[Nuclear starburst activity induced by elongated bulges in spiral galaxies]
{Nuclear starburst activity induced by elongated bulges in spiral galaxies}
\author[E. Kim et al.]{
Eunbin Kim$^{1}$\thanks{E-mail: ebkim@khu.ac.kr}, 
Sungsoo S. Kim$^{1,2}$, 
Yun-Young Choi$^{2}$,
Gwang-Ho Lee$^{3,4}$\thanks{KASI-Arizona Fellow}
\newauthor{Richard de Grijs$^{5,6,7}$, Myung Gyoon Lee$^{8}$ and Ho Seong Hwang$^{9}$}
 \\
$^{1}$School of Space Research, Kyung Hee University, Yongin, Gyeonggi 17104, Korea\\
$^{2}$Department of Astronomy \& Space Science, Kyung Hee University, 
Yongin, Gyeonggi 17104, Korea\\
$^{3}$Steward Observatory, University of Arizona, 933 North Cherry Avenue, Tucson, AZ 85721, USA\\
$^{4}$Korea Astronomy and Space Science Institute, Daejeon 305-348, Republic of Korea \\ 
$^{5}$Department of Physics and Astronomy, Macquarie University, Balaclava Road, Sydney, NSW 2109, Australia\\
$^{6}$Kavli Institute for Astronomy \& Astrophysics and Department of Astronomy,
Peking University, Yi He Yuan Lu 5, Hai Dian District,\\
 Beijing 100871, China\\
$^{7}$International Space Science Institute -- Beijing, 1 Nanertiao,
Zhongguancun, Hai Dian District, Beijing 100190, China\\
$^{8}$Department of Physics and Astronomy, Seoul National University,
1 Gwanak-ro, Gwanak-gu, Seoul 08826, Korea\\
$^{9}$Quantum Universe Center, Korea Institute for Advanced Study, 
  85 Hoegiro, Dongdaemun-gu, Seoul 02455, Korea\\
}
\date{Accepted XXX. Received YYY; in original form ZZZ}

\pubyear{2018}

\begin{document}
\label{firstpage}
\pagerange{\pageref{firstpage}--\pageref{lastpage}}
\maketitle
\begin{abstract}
We study the effects of bulge elongation on the star formation activity
   in the centres of spiral galaxies
  using the data from the Sloan Digital Sky Survey Data Release 7.
We construct a volume-limited sample of face-on spiral galaxies with
$M_r < -19.5$ mag at 0.02 $\leq z$< 0.055 by excluding barred galaxies, 
 where the aperture of the SDSS spectroscopic fibre covers
 the bulges of the galaxies.
We adopt the ellipticity of bulges measured by \cite{simard11}
 who performed two-dimensional bulge+disc decompositions using the SDSS images of galaxies, 
 and identify nuclear starbursts using the fibre specific star formation rates derived from the SDSS spectra. 
We find a statistically significant correlation between bulge elongation
  and nuclear starbursts in the sense that the fraction of
  nuclear starbursts increases with bulge elongation.
This correlation is more prominent for fainter and redder galaxies, 
 which exhibit higher ratios of elongated bulges. 
We find no significant environmental dependence of the correlation 
 between bulge elongation and nuclear starbursts.
These results suggest that non-axisymmetric bulges can efficiently 
  feed the gas into the centre of galaxies 
  to trigger nuclear starburst activity.

\end{abstract}




\section{Introduction\label{sec:intro}}

The development of observational techniques covering multiple wavelengths 
 has brought the detailed structures of galactic centres 
 such as nuclear discs, nuclear rings, and nuclear bars to light
 \citep{morris96,knapen02,comeron10,mazzuca11,alvarez15}. 
These central structures are believed to be the outcome 
 of the rearrangement of mass and energy 
 through triaxial potentials from bulges, bars, or ovals \citep{kormendy04}.
In order to explain the various structures in the central regions of galaxies, 
 migration of gaseous materials from the disc to the central region 
 must first be examined.
There are two main mechanisms driving the infall of gaseous materials.
First, external effects from interactions and mergers  
 of galaxies can affect the movement of gaseous materials \citep{haan13,medling14},
eventually triggering the star formation or nuclear activity in galaxies \citep{hwang11,hwang12}.
Second, internal processes can cause gaseous materials to fall in 
 via a non-axisymmetric potential which leads to secular evolution. 
This gas inflow is difficult to observe directly; however, 
 simulations have revealed that a non-axisymmetric mass distribution
 of triaxial structures can cause gas to lose angular momentum 
 and migrate to the central regions of galaxies \citep{shlosman90,athan94,combes01}.

To explain gas movement, early studies \citep{contopoulos80,binney91}
 suggested that gaseous materials move along the stable closed orbits (i.e., $x_1$ orbits) 
 elongated along the bar's major axis. 
The gaseous materials go through shocks and flow inwards, losing angular momentum, 
 to arrive on the $x_2$ orbits, which are elongated along the minor axis. 
Since the gaseous materials accumulated in the central regions 
 are used as fuel for high rates of starbursts, enhanced star formation rates 
 are a good tracer of recent gas inflow to the centre \citep{knapen95}.
These nuclear starbursts have strong H$\alpha$ emission lines and a size of 0.2 to 2 kpc 
 in the circumnuclear region \citep{kennicutt98}. 
They are occasionally observed as ring shapes, 
 i.e., nuclear rings \citep{knapen05,knapen06,comeron10}.
 
Bars have long been considered the primary non-axisymmetric structure of galaxies
 that can cause a triaxial potential, and
 have an influence on nuclear starbursts
 in galactic centres, as shown in many simulations.
 Hydrodynamic simulations show
 gas migration from the galactic disc to the central molecular zone 
 and enhanced star formation in the galactic centre \citep{kimss11,kimwt12,seo13,shin17}.
The relationships between bars and nuclear starbursts
 \citep{ho97,mulchaey97,knapen00,knapen06,ellison11,wang12,kim17}
  and between bars and nuclear rings \citep{martinet97,aguerri99,mazzuca08,comeron10}
have also been studied through many observations. 
Although two-thirds of galaxies have bars \citep{de91,mulchaey97,knapen00,eskridge00,lauri09},
 non-barred galaxies can also host ovals or triaxial bulges kinematically acting like bars.
Compared to bars, ovals have lower ellipticity \citep{lauri09}, 
 but more of the disc mass is involved in the non-axisymmetry \citep{kormendy04}. 
Since the dynamical evolution of ovals is similar to that of bars, 
 their kinematic effects would also be similar \citep{kormendy04}.

Triaxial bulges have been discovered in many spiral galaxies \citep{kormendy82,zaritsky86,bertola91}.
\cite{mendez08} showed that 80 per cent of bulges in non-barred lenticulars 
 and early to intermediate spiral galaxies are triaxial. 
The dynamics of triaxial bulges resemble those of bars, but
 they are different from those of elliptical galaxies \citep{kormendy82}. 
Non-axisymmetric potentials of triaxial bulges acting like bars are common,
 yet there are not enough studies to understand the relationship between bulges and nuclear starbursts.

In this paper, we study the effect of bulge elongation on nuclear starbursts of galaxies
  as a function of galaxy parameters that include 
  luminosity, colour, concentration and environments.
We use data from the Sloan Digital Sky Survey (SDSS; \citealt{york00}) Data Release 7,
  and only select non-barred galaxies to avoid mixed effects of bars 
  due to a bar and a bulge are morphologically and physically correlated \citep{kim15}.
We use 6,490 galaxies in the redshift range of 0.02 $\leq z<$ 0.055.
This allows us to statistically study the relationship between 
 the elongation of bulges and central starbursts on galaxy properties and the environment.
This paper is organised as follows. We describe our sample in Section 2, 
 and present our results on how triaxial bulges affect central starbursts in Section 3.
We summarise these results in the context of galaxy evolution in Section 4.

\section{SAMPLE SELECTION}

\subsection{A Volume-limited Sample of Non-barred Galaxies}\label{vol}
Our goal is to use a large sample of non-barred galaxies with 
  both photometric and spectroscopic data to study a possible correlation
  between bulge elongation and nuclear star formation activity.  
Thus, we use the SDSS, which is the largest survey 
 to explore galaxies and quasars with multi-colour images, and covers one-fourth of the sky. 
We selected a volume-limited sample of 33,391 galaxies 
 spanning the redshift range of 0.02 $\leq z $< 0.055 
 and the magnitude range of $M_r <  -19.5$+5log$\it{h}$.
Hereafter, we drop the +5log$\it{h}$ term in the absolute magnitude (${H_0}$ = 100 km s$^{-1}$ Mpc$^{-1}$ ). 
The bar classification for a large sample of SDSS galaxies is available in \cite{lee12}.
They provided a volume-limited sample for the same redshift and 
  $\it{r}$-band absolute magnitude ranges.
They visually inspected $\it{g}$ + $\it{r}$ + $\it{i}$ combined colour images, 
 and investigated the relationship 
 between the presence of bars and galaxy properties. 
This bar classification includes three different bar types: 
strong, weak and ambiguous bar types. 
Although visual inspection is still a reliable method 
  to identify internal features of galaxies and
  the classification of barred and non-barred galaxies 
  in this study agrees well with others \citep{nair10,huertas11}, 
  visual inspection might miss some galaxies 
  with very weak bars (Y. H. Lee et al. 2018, in preparation).
Our redshift and magnitude cuts allow us to study only 
   those galaxies which are large enough to show internal features. 
   We could also avoid some saturated galaxies or
   very faint galaxies using these cuts, 
  which can minimize any biases that could be introduced by visual inspection.

The catalogue of \cite{lee12} contains several physical parameters
 of galaxies including their morphology and photometry 
 drawn from the Korea Institute for Advanced Study 
 Value-Added Galaxy Catalogue (KIAS-VAGC; \citealt{choi10}), 
 which is based on the SDSS DR7 \citep{abazajian09}.
The galaxy morphology is determined using an automated 
 classification scheme of \cite{park05} and from additional visual classification.
From among the 33,391 galaxies in \cite{lee12}, 
  we selected only 10,830 late-type galaxies with an axis ratio of $b/a >$ 0.6.
The axis ratio condition is applied
 to reduce contamination by internal extinction effects and 
 selection bias because of inclination effects.
Because galaxies are expected to have random inclination angles on the sky,
  this condition does not introduce any bias in our sample selection.
Among the 10,830 late-type galaxies with an axis ratio of $b/a >$ 0.6,
  6,490 galaxies do not have bars.

\begin{table}
\centering \caption{Galaxy sample in this study}
\begin{tabular}{clc}
\hline
Step & Criteria & Number of galaxies \\
\hline
1 & Galaxies with 0.02 $\leq z<$ 0.055 & 33,391 \\
   &   \& $M_r < $ --19.5 mag &  \\
2 & Late types with $b/a >$ 0.6 & 10,830\\
3 & Non-barred galaxies & 6,490 \\
4 & Bulge+disc decomposition & 5,577\\
5 & Final sample  & 1,291\\
\hline
\end{tabular}
\label{tab:bar}
\end{table}

Physical parameters of galaxies used in this paper are provided by KIAS-VAGC: 
 absolute Petrosian magnitude $M_r^{0.1}$, ($\it{u-r})$$^{0.1}$  colour, 
 and inverse concentration index ($\it{c}_{in}$).
These parameters represent most major physical properties of galaxies
 that are related to star formation activity \citep{park09}. 
The rest-frame absolute magnitudes of individual galaxies were computed in fixed bandpasses, 
 shifted to $\it{z}$ = 0.1, using Galactic reddening corrections  \citep{schlegel98}. 
 $\it{K}$-corrections \citep{blanton03} and the mean evolution correction 
 \citep{tegmark04} were also applied.
Hereafter, the superscript 0.1, which represents the rest-frame at $\it{z}$ = 0.1, will be dropped.
The concentration index is defined as  $\it{c}_{in}$ = $R_{50}/R_{90}$ for an $\it{i}$-band image
 including seeing correction. 
$R_{50}$ and $R_{90}$  are the radii from the centre of a galaxy 
 containing 50$\%$, and 90$\%$  of the Petrosian flux, respectively.

\begin{figure}
\centering
\includegraphics{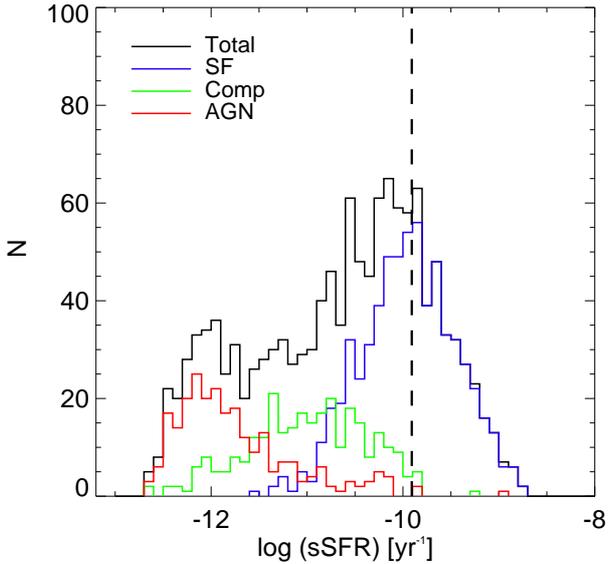}
\caption{ Distributions of total (black line), star forming (SF; blue line), composite (Comp; green line), 
  and AGN (red line) galaxies as a function of galactic sSFR. 
The vertical dashed line represents the peak value of the distribution.
}
\label{fig:ssfr}
\end{figure}
\begin{figure}
\centering
\includegraphics{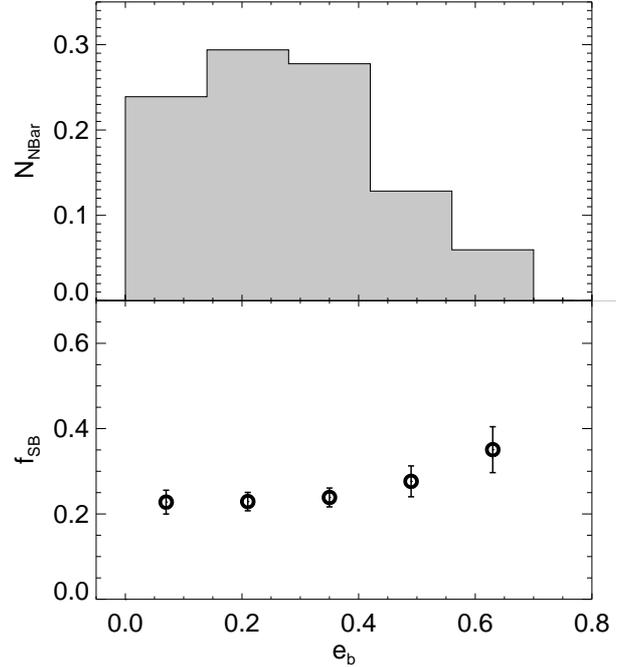}
\caption{ Distribution of non-barred galaxies (top) and fraction of central starburst galaxies (bottom)
 as a function of bulge ellipticity ($e_{\rm b}$).}
\label{fig:hist}
\end{figure}

\subsection{Identification of bulge+disc systems using the two-dimensional decomposition of Simard et al. (2011)}

We cross-matched the 6,490 non-barred galaxies with the galaxies in
  the catalogue of \cite{simard11} who performed two-dimensional 
  bulge+disc decompositions using the SDSS $\it g$ and $\it r$ images
  for 1,123,718 galaxies in the SDSS DR7. 
They used the GIM2D software package \citep{simard02} for
  the decomposition of galaxy images.
They provide the results from three different galaxy fitting models which
  include a pure S\'ersic model and 
  S\'ersic bulge + disc models with free S\'ersic index ($n_b$) or 
  with fixed $n_b$ (i.e. $n_b$ = 4).
We adopt the bulge parameters (e.g. ellipiticy, effective radius)
  from the most general fitting case 
  (i.e. S\'ersic bulge + disc model with free S\'ersic index).
\cite{kim16} showed that 
  the results of \cite{simard11} 
  based on the S\'ersic bulge + disc model with free S\'ersic index
  agree well with their results using GALFIT \citep{peng02}.
Although the galaxy sample of \cite{simard11} is also from SDSS DR7,
  the sample selection is not exactly the same as the one
  for our parent galaxy sample (i.e. KIAS-VAGC).
The KIAS-VAGC also contains some galaxies 
  with measured redshifts from the literature, 
  not included in the catalogue of \cite{simard11}.
Among the 6,490 galaxies in our sample,
  there are 5,577 galaxies with measured bulge parameters.
 
To select the bulge+disc systems with reasonable profile fitting results,
  we apply the following selection criteria 
  by combining the conditions recommended by \cite{simard11};
1) the $F$-test probability ($P_{pS}$) that
 a bulge+disc model is not required  compared  to a pure S\'ersic model should be 
 equal to or higher than 0.32 (i.e. $P_{pS} \ge$ 0.32),
2) the effective radius of a bulge should be larger than two pixels\footnote{The pixel size 
  and the typical seeing of the SDSS observations are
   0.396\arcsec and 1.43\arcsec respectively in the $\it{r}$-band. When we conservatively select the galaxies with $r_{\rm eff,b}$ > 5 pixels instead of $r_{\rm eff,b}$ > 2 pixels, our findings in the present paper remain largely unaltered.}
   (i.e. ${r_{\rm eff,b}}>$ 2 pixels).
3) the bulge fraction ($B/T$) should be larger than 0.2, 
4) the disc inclination angle measured in \cite{simard11} should be equal to or less than 53 degree 
  that corresponds to axis ratio 0.6 as applied to the KIAS-VAGC in Section \ref{vol}, and
5) the S\'ersic index ($n_b$) should be larger than 0.5 and 
  smaller than 8: i.e. 0.5 $< n_b <$ 8;
  the galaxies with $n_b$ = 0.5 or 8 are those 
  with nuclear sources, off-center components, etc.  
The galaxies with large errors in measured bulge ellipticity
  are already removed by these criteria, thus
  we do not include the condition for the bulge ellipticity error in these criteria;
  the mean error of measured bulge ellipticity 
  in the final sample is $e_{\rm b}$ is 0.022 $\pm$ 0.002.
The use of different criteria can change the number of sample galaxies, 
  but does not change our main conclusion.
We also visually inspected the galaxies to remove 53 problematic cases
 (e.g. merging galaxies, irregular galaxies, and contaminated by bright stars). 
We are left with a final sample of 1,291 galaxies.
Table \ref{tab:bar} summarises the changes of galaxy numbers in these steps.

 \subsection{Identification of Galaxies with Central Starbursts}

In the redshift range of 0.02 $\leq z <$ 0.055, 
  the physical size of the SDSS fibre radius (1.5$^{\prime\prime}$) corresponds 
  to 0.44 to 1.18 kpc.
Effective radii of galactic bulges are up to $\approx$ 3 kpc 
 in the similar redshift range \citep{gadotti09,fisher10}.
Thus the size of the SDSS fibre radius is similar to or smaller than 
 the size of the galaxy bulges. 
For this reason, we assume that the star formation rates derived from the fibres
 represent those from the central regions (bulges) of the galaxies.
The specific SFR (sSFR) values are taken from the Max Planck Institute for Astrophysics and
 Johns Hopkins University (MPA/JHU) DR8 catalogue \citep{brinchmann04,kauffmann03}.
The fibre (hereafter, nuclear) sSFRs distribution of our sample is shown in Fig. \ref{fig:ssfr}.
Objects are classified as star-forming galaxies, composite galaxies, 
  or active galactic nuclei (AGN)
  using the Baldwin-Phillips-Terlevich (BPT) diagram \citep{baldwin81,kewley06}. 
We fit double Gaussians to the log(sSFR) distribution of our sample,
 and find that the larger of the two Gaussian peaks has a value of log(sSFR)$= -9.98$ [yr$^{-1}$].
We define galaxies with log(sSFR)$\ge$ -- 9.98 [yr$^{-1}$] as central starburst galaxies
  which include star forming galaxies of higher log(sSFR) than the median value.
Fig. \ref{fig:ssfr} shows that AGNs are clearly separated 
 from our starburst galaxies in the sSFR dimension.
We will discuss the degree of central activity of a certain galaxy group 
 in terms of the fraction of nuclear starburst galaxies, $f_{\rm SB}$, in the following sections.

\begin{figure*}
\centering
\includegraphics{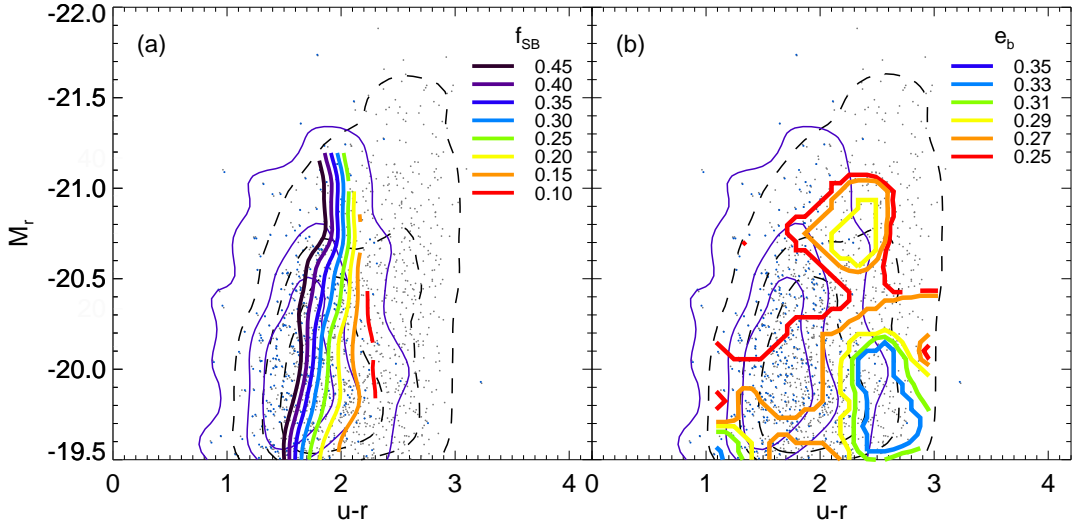}
\caption{(a) Starburst fraction ($f_{\rm SB}$)  and (b) median bulge ellipticity ($e_{\rm b}$) contours
 in $\it{u-r}$ colour vs. $M_r$.  Blue and gray dots represent SB and non-SB galaxies. 
Solid blue and dashed black contours represent the 0.5$\sigma$, 1$\sigma$ and 2$\sigma$ 
 number densities of the blue and total galaxy samples from the inside to the outside.}
\label{fig:colormag}
\end{figure*}


\section{Results}

\subsection{Correlation between bulge elongation and nuclear star formation activity in galaxies}

Here we examine the correlation between 
  bulge elongation and nuclear star formation activity. 
We first show a fraction of non-barred galaxies with nuclear starburst activity ($f_{\rm SB}$)
 as a function of bulge ellipticity (${e_{\rm b}}$) in the bottom panel of Fig. \ref{fig:hist}.
The $f_{\rm SB}$  increases with $e_{\rm b}$. 
The number distribution in the top panel shows that 
 the peak of the distribution is around $e_{\rm b}=0.3$, 
 and the number of galaxies decreases at $e_{\rm b} >$ 0.4. 
We use several statistical tools to examine the significance of 
  the correlation between bulge ellipticity and starburst fraction. 
 The Spearman correlation test between the two gives 
   a correlation coefficient ${\rho_{\rm corr}}$ =  1.0 and 
   the probability of obtaining the correlation by chance of ${p_{\rm corr}}<$ 0.001, 
   suggesting a significant correlation.
We also apply the Kolmogorov-Smirnov (K-S) test and
   the Anderson-Darling (A-D) k-sample test directly
    to the distributions of sSFRs of galaxies
     (not the fraction) for two subsamples divided by bulge ellipticity (i.e. ${e_{\rm b}} >$ 0.4 and ${e_{\rm b}}$ $\leq$ 0.4).
We could reject the hypothesis that the sSFR distributions
   of the two samples are extracted from the same parent population
    with a confidence level of 98$\%$.
This confirms a significant difference in the star formation activity
   between the two subsamples with different bulge ellipticities.

Fig. \ref{fig:colormag}(a) shows that $f_{\rm SB}$ 
  depends more strongly on the $\it{u-r}$ colour than on $M_r$. 
Here, note that fixing the  $\it{u-r}$ colour  is needed to carefully
 investigate the relationship between bulge elongation and nuclear star formation.
We divided samples into relatively blue and red at $\it{u-r} \approx$ 1.8 mag,
  which is similar to the peak of $\it{u-r}$ colour of spiral galaxies separating morphological type Sb vs. SC/Irr
 (See Fig. 7 in \citealt{strateva01,badlry04}).
For blue galaxies with $\it{u-r}<$ 1.8 mag, bright galaxies have higher $f_{\rm SB}$ at fixed  colour.
These galaxies might be the result of gas-rich major mergers 
 that show high star formation rates during late stages of merging \citep{kennicutt12,vandokkum05}.
Fig. \ref{fig:colormag}(b) shows that bulges are more elongated 
 as galaxies become redder and fainter. 
$e_{\rm b}$ has low values 
 when galaxies become brighter than $M_r \approx -20.4$ mag. 
 This value is similar to a characteristic luminosity in the $\it{r}$-band,
 $M_\star \approx$ $-20.4$ mag, of the SDSS sample \citep{blanton03}. 
Galaxies brighter than the characteristic luminosity show that
 the number density, stellar mass, gas contents and other parameters
 dramatically change compared to galaxies with lower luminosities than
 the characteristic luminosity \citep{blanton09}.
Since $f_{\rm SB}$  and $e_{\rm b}$  are intricately correlated with the magnitude and colour of host galaxies, 
 we divided the samples into bright or faint galaxies, and blue or red galaxies 
 in order to separate the intricate relationships between the parameters in our subsequent analysis. 
The  $f_{\rm SB}$  and $e_{\rm b}$ contours indicate
  the fraction and the median value at each point, respectively. 
We obtain the contours by dividing each panel into 60 by 60 bins
  and by applying the spline kernel method to extract smoothed distributions.
The contours represent 2$\sigma$ level, and
  the uncertainties for $f_{\rm SB}$ and $e_{\rm b}$ 
  are calculated by 1000 times resampling bootstrap method.
\begin{figure*}
\centering
\includegraphics{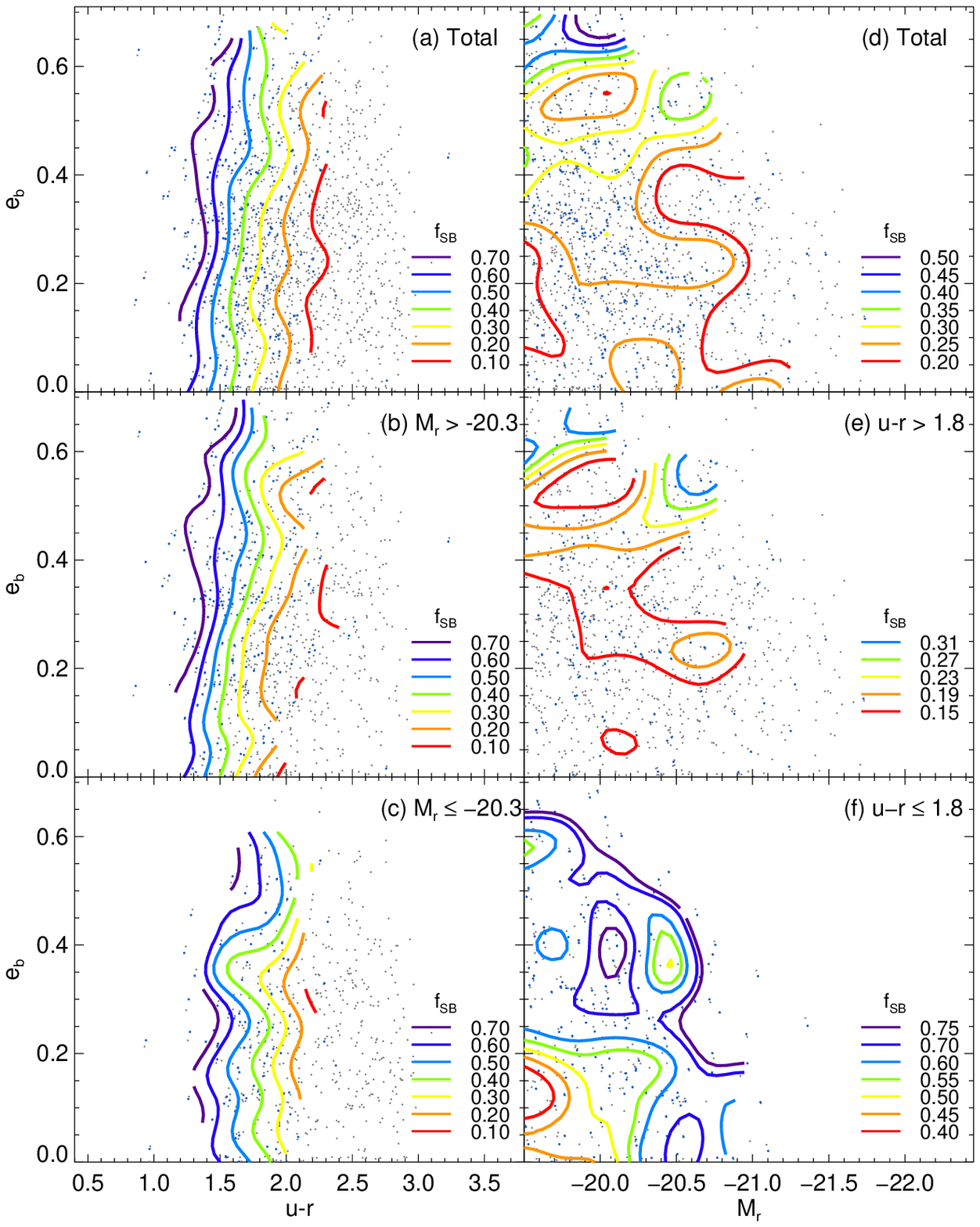}
\caption{Starburst fraction ($f_{\rm SB}$) contours in (left) bulge ellipticity vs. $\it{u-r}$ colour and (right) bulge ellipticity vs. $M_r$. In the left column, samples are (a) the whole sample, (b) bright galaxies with $M_r <$ -20.3 mag, and (c) faint galaxies $M_r >$ -20.3 mag. In right column, samples are (d) the whole sample,  (e) red galaxies with $\it{u-r}>$ 1.8 mag, and (f) blue galaxies with $\it{u-r} <$ 1.8 mag. Blue and gray dots represent SB and non-SB galaxies.}
\label{fig:tot}
\end{figure*}

\subsection{The Effects of Absolute magnitudes ($M_r$) and $\it{u-r}$ colours 
  on the Correlation between bulge elongation and nuclear star formation activity}

We compare the dependence of $f_{\rm SB}$ on several physical parameters 
 when galaxies are classified as bright or faint, and blue or red.
Fig. \ref{fig:tot} shows $f_{\rm SB}$ of the galaxies
 in the planes of $e_{\rm b}$ versus $\it{u-r}$ colour and of $e_{\rm b}$ versus $M_r$.
The left panels are for the whole galaxy sample, realtively
  faint ($M_r > -20.3$ mag ) and bright ($M_r \leq -20.3$ mag ) galaxies.
In Fig. \ref{fig:tot}(a), $f_{\rm SB}$  of all galaxies strongly depends on colour
 and increases from 10$\%$ to 70$\%$ as the colour becomes bluer.
This stronger dependence of $f_{\rm SB}$ on colour
  than on $e_{\rm b}$ remains similar even when we divide
  the galaxies into two subsamples based on luminosity (i.e. middle and bottom panels).

The right panels show the dependence of $f_{\rm SB}$
  on $e_{\rm b}$ and $M_r$ for 
 total, relatively red ($\it{u-r} >$ 1.8 mag) and blue galaxies  ($\it{u-r} \leq$ 1.8 mag). 
Fig. \ref{fig:tot}(d) shows that it is difficult to separate the effects
  of $e_{\rm b}$ and $M_r$ on $f_{\rm SB}$ because of noisy contours.
When the samples are divided by $\it{u - r}$ colour,
 the dependence of $f_{\rm SB}$ on $e_{\rm b}$ and on $M_r$
 become more well-defined.
For example, Fig. \ref{fig:tot}(e) based on the sample of red galaxies 
  containing relatively low gas amounts
  shows that the contours are more horizontal at $e_{\rm b}>$ 0.2. 
This suggests that $f_{\rm SB}$ is correlated better with $e_{\rm b}$ 
  than with $M_r$.
Fig. \ref{fig:tot}(f) shows the dependence of 
  $f_{\rm SB}$ on $e_{\rm b}$ in blue galaxies 
 which are expected to have relatively large amounts of gas. 
The comparison of panels (e) and (f) suggests that
  $f_{\rm SB}$ is generally higher in blue galaxies than in red galaxies. 
The $f_{\rm SB}$ increases as $e_{\rm b}$ increases when galaxies are relatively faint.
The $f_{\rm SB}$ also increases as $M_r$ brightens at a given $e_{\rm b}$ 
  when  $e_{\rm b}$ is smaller than 0.2.
It also suggests that
 the bulge elongation effect on $f_{\rm SB}$ is slightly more prominent 
 in red galaxies than in blue galaxies.

\begin{figure*}
\centering
\includegraphics{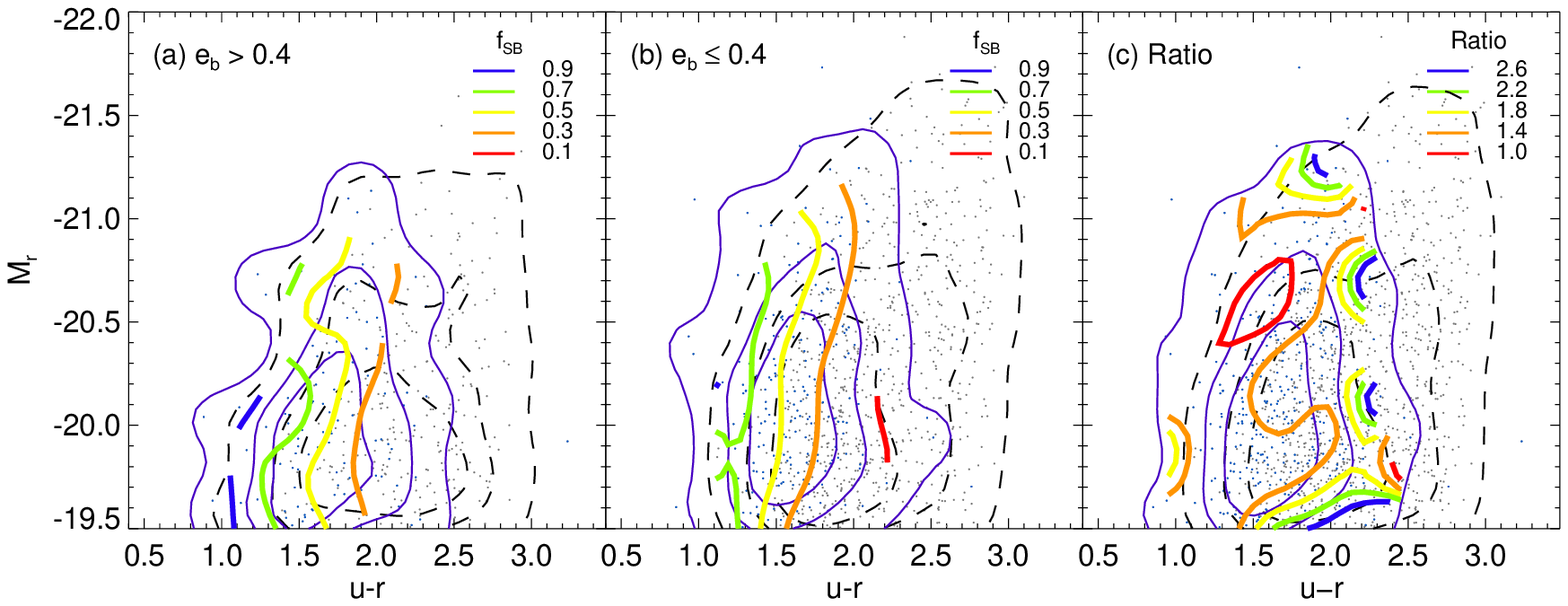}
\caption{Starburst fraction ($f_{\rm SB}$) contours of galaxies with (a) elongated bulges ($e_{\rm b}>$ 0.3), 
(b) round bulges ($e_{\rm b}<$ 0.3), and (c) the ratio of (a) and (b). 
Solid blue and dashed black contours represent the 20$\%$, 50$\%$, and 90$\%$ 
number densities of the blue and total galaxy samples from the inside to the outside.
Blue and gray dots represent SB and non-SB galaxies, respectively. \label{fig:ratio}}
\end{figure*}

The $f_{\rm SB}$ of galaxies that have relatively
 elongated bulges ($e_{\rm b} >$ 0.4) and rounded bulges ($e_{\rm b} <$ 0.4), 
 and their ratios are shown in Fig. \ref{fig:ratio}. 
Galaxies with elongated bulges are a few in bright region, and
  the contours of $f_{\rm SB}$ for galaxies with elongated bulges 
  are shifted to redder colours compared with those with rounded bulges in Fig. \ref{fig:ratio} (a) and (b). 
The ratio of $f_{\rm SB}$ of galaxies with elongated and rounded bulges
  is shown in Fig. \ref{fig:ratio} (c).
There is a trend that the ratio increases as $\it{u-r}$ becomes redder, and
 the ratio for faint galaxies is higher compared to that for bright galaxies.
This implies that these bright galaxies that have grown through mergers 
 have low $e_{\rm b}$, and weak correlation with $e_{\rm b}$.
The $e_{\rm b}$ effects on the $f_{\rm SB}$ of faint galaxies depending on colour 
 can be more specifically explained based on the bulge dominance in Section 3.3.


\subsection{The Effects of Mass Concentration on the Correlation between bulge elongation and nuclear star formation activity}

\begin{figure}
\centering
\includegraphics{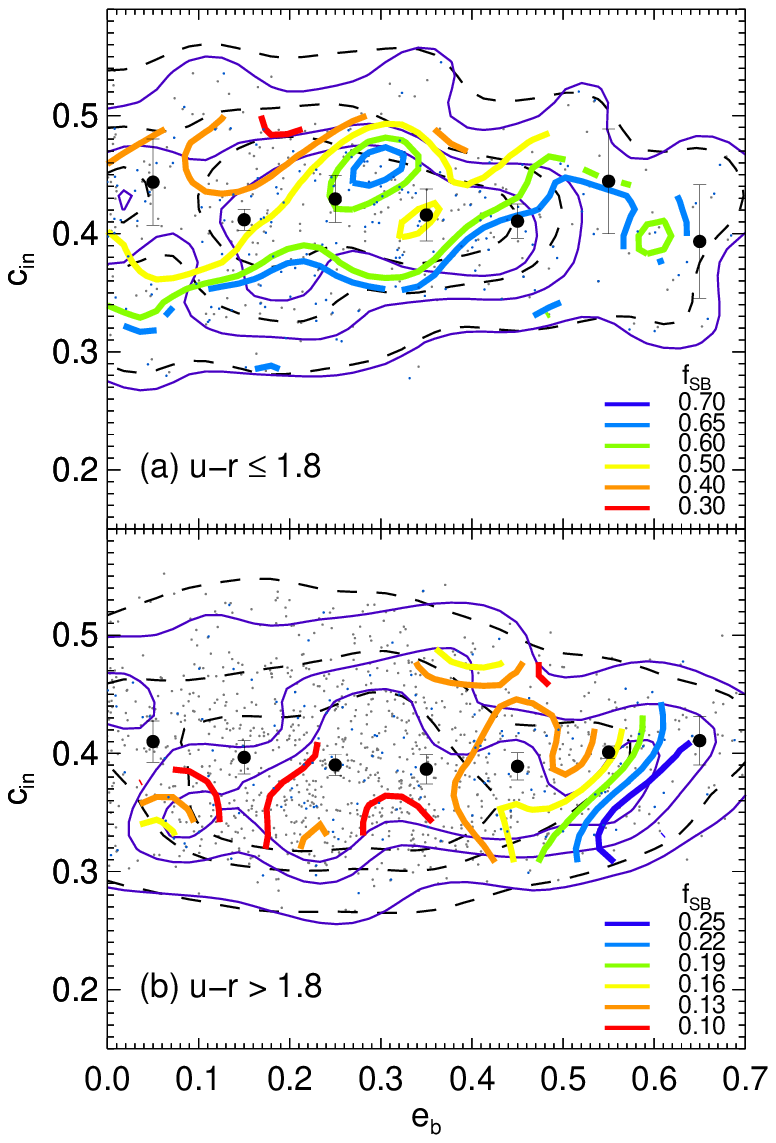}
\caption{Starburst fractions ($f_{\rm SB}$) contours of (a) blue ($\it{u-r} <$ 1.8 mag) 
 and (b) red ($\it{u-r} <$ 1.8 mag) galaxies as a function of the concentration index $c_{\rm in}$ and bulge ellipticity $e_{\rm b}$.
Black circles represent the median of $c_{\rm in}$ in each bin. Blue and gray dots are SB galaxies and non-SB galaxies, respectively.}
\label{fig:cin}
\end{figure}

In the previous sections, we found that the role of bulge ellipticity 
 in nuclear starbursts depends on galaxy properties.
Here, we examine how the role of bulge ellipticity differs
 according to the light concentration (bulge dominance),
 in addition to the colours and luminosities of galaxies.
The luminosity distribution of bulges has a relationship 
 with the host galaxy's morphological type \citep{andre95,graham00} and 
 separates galaxies with different star formation histories \citep{choi07}. 
The bulge dominance can be determined using the bulge-to-total flux ratio, 
 the concentration index, or the inverse concentration index \citep{shimasaku01,park05,gadotti09}.
We use the inverse concentration index \citep{shimasaku01,park05},
  $c_{\rm in}$, as a bulge dominance indicator.
Fig. \ref{fig:cin} shows that for blue galaxies 
  $f_{\rm SB}$ is tightly correlated with $c_{\rm in}$ and
  the galaxies with high $e_{\rm b}$ on average have high  $f_{\rm SB}$.
For red galaxies, 
  $f_{\rm SB}$ is not related to  $c_{\rm in}$ as much as for blue galaxies, 
  but very clearly the galaxies have higher  $f_{\rm SB}$ as $e_{\rm b}$ increases.
 
 \subsection{ Environmental Effects on the Correlation between bulge elongation and nuclear star formation activity}

The environments of galaxies affect not only the internal structure of galaxies, 
  but also various activities in galaxies \citep{park07,hwang10,lee18}.
\cite{lackner13} found two different environmental effects: 
  relatively high density environments affect morphological transformation, 
  while low density environments contribute to star formation quenching.
Bulges also have grown through galaxy evolution 
  driven by environmental or bar effects \citep{mendez08}.
Thus, we need to separate the environmental effects that
 can influence bulges and nuclear star formation.
 
We adopt two types of environmental parameters: a background mass density $\rho_{20}$ as a large-scale environment
  and a projected distance to the nearest neighbour galaxy R$_{\rm n}$ as a small-scale environment.
For the large-scale environment, 
  we select galaxies from intermediate density regions
  to avoid extreme environments such as clusters or void regions.
To exclude the effects of neighbouring galaxies,
  we select isolated galaxies 
  that are placed far away (3 times their virial radius from neighbouring galaxies)
  and investigate the correlation of $e_{\rm b}$ and $f_{\rm SB}$ for these galaxies. 
For comparison, we present the results of galaxies that 
 have neighbouring galaxies at relatively close distances (i.e. interacting galaxies).

The large-scale background density  \citep{park09}, the mass density,
  is determined by using 20 neighbouring galaxies over a few Mpc scale. 
The small-scale environmental parameter is the normalised distance to the nearest neighbour galaxy. 
The background density at a given location of a galaxy is obtained through Eq. (1),
 
\begin{equation}
\rho_{20}(x)/\bar{\rho}=\sum_{i=1}^{20}{\gamma_{i}}L_{i}W_{i}(|x_{i}-x|)/\bar{\rho},
\end{equation}

where $\it{x}$ is the location of the target galaxy.
Parameters  $\gamma_i$, $L_i$, and $\bar{\rho}$ are the mass-to-light ratio, 
 the $\it{r}$-band luminosity of the nearest 20 galaxies brighter than $M_r = -19.5$ mag, 
 and the mean density of the universe, respectively. 
The smoothing filter function, the spline-kernel weight, $W$,
 and other detailed information are described in \cite{park08} and \cite{park09}.
Second, we consider the distance to the nearest galaxy from the target galaxy, 
 normalised by the virial radius of the nearest galaxy. 
This parameter is expressed as $R_{\rm n}/r_{{\rm vir},n}$.
The virial radius of a galaxy $r_{{\rm vir},n}$ is defined as the projected radius in which
 the mean mass density is 740 times the mean density of the universe. That is,

\begin{equation}
r_{{\rm vir},n}=(3\gamma L/ 4\pi / 740\bar{\rho})^{1/3}h^{-1} {\rm Mpc}.
\end{equation}

\begin{figure}
\centering
\includegraphics{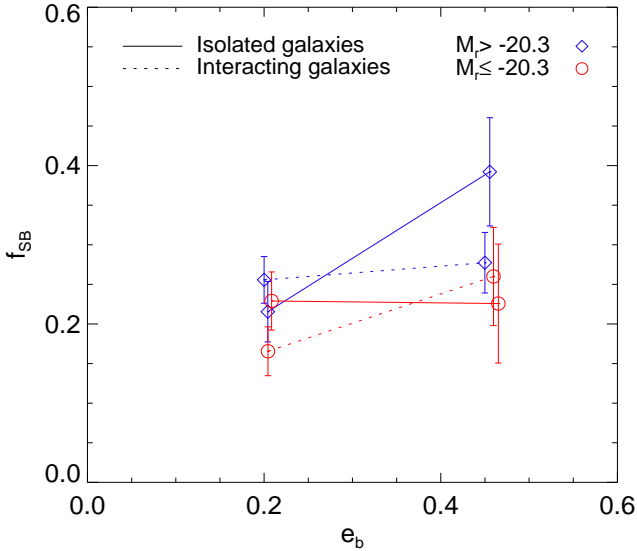}
\caption{Starburst fractions ($f_{\rm SB}$) of galaxies as a function of $e_{\rm b}$. 
 Solid and dotted lines represent isolated and interacting galaxies, respectively.
 Red and blue represent bright and faint galaxies, respectively.}
\label{fig:rp}
\end{figure}

See Section 2.3 in \cite{park08} and Section 2.3 in \cite{park09} for detailed descriptions. 
Due to $\rho_{20}/\bar{\rho}$ at the locations in massive clusters is over 50 \citep{park09}, 
  we constrain the sample considering the environmental effects and 
  the number of galaxies based on the following criteria. 
We select galaxies at intermediate density 
  (1 $<\rho_{20}/\bar{\rho}<$ 30), then divide them into
  isolated ($R_n > 3 r_{{\rm vir},n}$) and interacting ($R_n < 0.7r_{{\rm vir},n}$) galaxies.
As we explained above, $R_n/r_{vir,n}$ is the parameter used to perceive 
 how far the host galaxies are from their neighbouring galaxies' virial radius.
When $R_ n/r_{{\rm vir},n} <$ 1, a host galaxy is located 
  within the virial radius of its nearest neighbour galaxy. 
Fig. \ref{fig:rp} shows that $f_{\rm SB}$
  of each galaxy sample divided by environment and luminosity
  generally increases with $e_{\rm b}$ despite large error bars,
   but the trend between $f_{\rm SB}$ and $e_{\rm b}$ 
  does not depend significantly on galaxy environment.
  

\section{Summary}
We use bulge ellipticity and physical parameters of galaxies 
 to examine the correlation between the elongation of bulge and the nuclear starbursts.
To better understand the bulge effect, we only use non-barred galaxies.
Because the fraction of galaxies with nuclear starburst activity and 
  the bulge elongation are affected by $M_r$ and $\it{u-r}$ colour, respectively, 
  we examine the correlation between the nuclear starburst fraction 
  and bulge ellipticity by fixing galaxy luminosity and colour. 
We find that $f_{\rm SB}$ generally increases in galaxies with larger bulge elongation,
 and this correlation is more prominent in faint and red galaxies.
 The major results are as follows.

\begin{enumerate}

\item The nuclear starburst fraction ($f_{\rm SB}$) is 
  more strongly dependent on $\it{u-r}$ colour than on $M_r$,
  which can indicate the importance of the amount of gas in star formation activity.
The bulge ellipticity increases as galaxies become fainter and redder.

\item The effects of elongated bulges on the nuclear starburst activity are more pronounced
 in fainter  (less massive) and redder galaxies (little gas supply).
This can suggest that a secular process associated with elongated bulges plays an important part in
 less massive galaxies with little cold gas reservoir, which seem to be in relatively sterile
 conditions for triggering nuclear starbursts.

\item $f_{\rm SB}$ also strongly depends on $c_{\rm in}$ for blue galaxies, but
  the $c_{\rm in}$ dependence of $f_{\rm SB}$ becomes weak and  
  the $e_{\rm b}$ dependence becomes strong for red galaxies. 
  
\item $f_{\rm SB}$ generally increases with $e_{\rm b}$   
  even when separating the galaxy environment.
However, this dependence does not differ much depending on galaxy environment.

\end{enumerate}
Our results suggest that non-axisymmetric bulge can feed the gas into 
  the centre of galaxies to trigger nuclear starburst activity. 
To better understand the correlation between star formation activity and bulge elongation,
  two-dimensional spectroscopic data along with information 
  on the amount of gas in these galaxies  will be very helpful.

\section*{Acknowledgements}

We thank the anonymous referee for insightful comments.
SSK and EK were supported by a National Research Foundation grant funded by
the Ministry of Science, ICT and Future Planning of Korea (NRF-2014R1A2A1A11052367).
GHL is supported by a KASI-Arizona Fellowship.
RdG was partially supported by the National Natural Science
Foundation of China through grants U1631102, 11373010, and 11633005.
RdG also acknowledges support from the National Key Research 
and Development Program of China through grant 2017YFA0402702.
MGL was supported by a grant from the National Research Foundation
(NRF) of Korea, funded by the Korean Government (NRF-2017R1A2B4004632).

Funding for the SDSS and SDSS-II has been provided by the Alfred P. Sloan Foundation, the Participating Institutions, the National Science Foundation, the U.S. Department of Energy, the National Aeronautics and Space Administration, the Japanese Monbukagakusho, the Max Planck Society, and the Higher Education Funding Council for England. The SDSS Web Site is \url{http://www.sdss.org/}.

The SDSS is managed by the Astrophysical Research Consortium for the Participating Institutions. The Participating Institutions are the American Museum of Natural History, Astrophysical Institute Potsdam, University of Basel, University of Cambridge, Case Western Reserve University, University of Chicago, Drexel University, Fermilab, the Institute for Advanced Study, the Japan Participation Group, Johns Hopkins University, the Joint Institute for Nuclear Astrophysics, the Kavli Institute for Particle Astrophysics and Cosmology, the Korean Scientist Group, the Chinese Academy of Sciences (LAMOST), Los Alamos National Laboratory, the Max-Planck-Institute for Astronomy (MPIA), the Max-Planck-Institute for Astrophysics (MPA), New Mexico State University, Ohio State University, University of Pittsburgh, University of Portsmouth, Princeton University, the United States Naval Observatory, and the University of Washington.





\bsp	
\label{lastpage}
\end{document}